\documentclass{emulateapj}
\usepackage{latexsym}
\def\be{\begin{equation}}
\def\ee{\end{equation}}
\def\bea{\begin{eqnarray}}
\def\eea{\end{eqnarray}}
\slugcomment{Accepted for publication in ApJ}
\shorttitle{High energy photon and neutrino emission from GRBs} 
\shortauthors{Fan, Zhang, \& Wei}

\begin{document}
\title{Early photon-shock interaction in a stellar wind: a sub-GeV photon
flash and high energy neutrino emission from long GRBs}  

\author{Y. Z. Fan$^{1,2,3}$, Bing Zhang$^{1}$ and D. M. Wei$^{2,3}$}

\affil{$^1$ Dept. of Physics, University of Nevada, Las Vegas, NV
89154, USA.\\ $^2$ Purple Mountain Observatory, Chinese Academy of
Science, Nanjing 210008, China.\\ $^3$ National Astronomical
Observatories, Chinese Academy of Sciences, Beijing, 100012, China.\\}

\begin{abstract}

For gamma-ray bursts (GRBs) born in a stellar wind, as the reverse
shock crosses the ejecta, usually the shocked regions are still
precipitated by the prompt MeV $\gamma-$ray emission. Because of the
tight overlapping of the MeV photon flow with the shocked regions, the
optical depth for the GeV photons produced in the shocks is very
large.  These high energy photons are absorbed by the MeV photon flow
and generate relativistic $e^\pm$ pairs. These pairs re-scatter the soft
X-ray photons from the forward shock as well as the prompt
$\gamma-$ray photons and power detectable high energy emission,
significant part of which is in the sub-GeV energy range.  Since the
total energy contained in the forward shock region and the reverse shock
region are comparable, the predicted sub-GeV emission is independent
on whether the GRB ejecta are magnetized (in which case the reverse
shock IC and synchrotron self-Compton emission is suppressed).  As a
result, a sub-GeV flash is a generic signature for the GRB wind model,
and it should be typically detectable by the future {\em Gamma-Ray
Large Area Telescope} (GLAST).  Overlapping also influence neutrino
emission. Besides the 
$10^{15} \sim 10^{17}$ eV neutrino emission powered by the interaction
of the shock accelerated protons with the synchrotron photons in both
the forward and reverse shock regions, there comes another $10^{14}$eV
neutrino emission component powered by protons interacting with the
MeV photon flow. This last component has a similar spectrum to the one
generated in the internal shock phase, but the typical energy is
slightly lower.

\end{abstract}


\section{Introduction}
The leading model for long gamma-ray bursts (GRBs) invokes
a relativistic jet emerging from a collapsar 
(e.g. Woosley 1993; Paczy\'{n}ski 1998). It has been widely
expected that GRBs associated with supernovae should occur in a
pre-burst stellar wind enviroment with particle number density
$n\propto R^{-2}$ (e.g. Dai \& Lu 1998; M\'{e}sz\'{a}ros, Rees \&
Wijers 1998; Chevalier \& Li 2000). As the jet expands into a
dense steller wind, a forward shock (FS) and a reverse shock (RS)
form (Chevalier \& Li 2000; Dai \& Lu 2001, hereafter DL01), whose
synchrotron emission peaks in the soft X-ray band and in the
ultraviolet to optical band, respectively.
Photomeson interaction in the RS region also results in neutrino
emission in the energy range of $3\times 10^{15}$ eV - $3 \times
10^{17}$ eV (DL01).

However, until very recently, in nearly all the works on
RS emission, possible overlapping of the initial prompt
$\gamma-$rays emission (MeV photon flow) with the RS region has been
ignored. 
Beloborodov (2005) pointed out that when overlapping is important, the
RS emission would be suppressed significantly since the electron
cooling is dominated by inverse Compton (IC) scattering off the MeV
photon flow. Correspondingly, the upscattering would power a strong
GeV-TeV photon flash. Fan et al. (2005a) suggested that 
when overlapping is substantial, similar processes are also relevant
to the early FS emission.

In this work, we show that overlapping is a common feature for GRBs
born in a steller wind (\S{\ref{overlapping}), so that the early
afterglow wind model needs substantial revision. We then systematically
study the consequence of this overlapping effect in the wind model. 
In \S{\ref{Sub-GeV}} we show that such an early photon-shock
interaction inevitably leads to the prediction of a sub-GeV 
photon flash in the wind model, which is generally detectable by the
{\em Gamma-Ray Large Area Telescope} (GLAST). In \S{\ref{neutrinos}},
we discuss various high energy neutrino emission processes in the early
afterglow stage as the result of photon-shock interaction.
We summarize our conclusions in \S{\ref{DisCon}} with some discussions.

\section{Tight overlapping of the MeV photon flow with the shocked
regions}\label{overlapping} 

The overlapping of the MeV photon flow with the fireball ejecta is
important if the RS crossing radius satisfies
\be
R_\times \ll 2\Gamma_0^2 \Delta,
\ee
where $\Gamma_0$ is the initial Lorentz factor of the ejecta (at the
internal shock phase), $\Delta=cT_{\rm 90}/(1+z)$, $T_{90}\sim 20{\rm
s}$ is the observed duration of the long GRB, and $z\sim 1$ is the
redshift. The shock crossing radius $R_\times$ can be written in a
general form
\begin{equation}
R_\times = {\rm max} (R_\gamma, \Gamma^2 \Delta),
\label{R_times}
\end{equation}
where $R_\gamma$ is the radius where the mass of the medium collected
by the fireball is equal to $1/\Gamma_0$ of the fireball mass (which
corresponds to $\Gamma \sim \Gamma_0 /2$), and $\Gamma$ is the Lorentz
factor of the shocked ejecta. In equation (\ref{R_times}), the first
term dominates if the RS is non-relativistic, while the second term
dominates if the RS is relativistic.  For the ISM case with a typical
number density $n_{\rm ISM}\sim 1{\rm cm^{-3}}$, the RS is
non-relativistic and one has $R_\times ({\rm ISM})=R_\gamma \approx
8.8\times 10^{16}E_{53.6}^{1/3}n_{\rm
ISM,0}^{-1/3}\Gamma_{0,2.5}^{-2/3}$, which is usually larger than (or
at least comparable to) $2\Gamma_0^2
\Delta= 5.4\times 10^{16}{\rm
cm}~\Gamma_{0,2.5}^2\Delta_{11.5}$. Throughout this paper, we adopt
the convention $Q_{\rm x}=Q/10^{\rm x}$ to express physical
parameters in cgs units. So, in the ISM case, the initial
$\gamma-$rays can not change the RS emission significantly. In the
stellar wind case, on the other hand, one has $n=3.0\times
10^{35}A_*R^{-2}{\rm cm^{-3}}$, where
$A_*=(\dot{M}/10^{-5}M_{\odot}~{\rm yr^{-1}})({\rm v}_{\rm w}/10^8{\rm
cm~s^{-1}})^{-1}$, $\dot{M}$ is the mass loss rate of the progenitor,
and ${\rm v}_{\rm w}$ is the wind velocity (Chevalier \& Li
2000). With such a dense medium, the RS is relativistic (Chevalier \&
Li 2000; DL01): $\Gamma\approx
\bar{\zeta}^{1/4}\Gamma_0^{1/2}/\sqrt{2}$, where $\bar{\zeta}\approx
2600E_{53.6}/(A_*\Delta_{11.5}\Gamma_{0,2.5}^2)$ is the ratio
of proper mass density of the unshocked ejecta and the medium density,
$E \sim 4\times 10^{53}{\rm ergs}$ is the isotropic energy of
the ejecta, which is calculated by adopting $T_{90}=20$s, $z=1$, and
by assuming that the observed $\gamma-$ray luminosity is $L_{\gamma}
\simeq 10^{52}{\rm ergs~{\rm s}^{-1}}$ and that the radiation
efficiency of the internal shocks is $\eta \simeq 0.2$. The total wind
luminosity is therefore $L_{\rm tot}=L_\gamma / \eta = 5\times 10^{52}{\rm
ergs~s^{-1}}$, and the afterglow kinetic energy luminosity is $L
\simeq L_{\rm tot}(1-\eta) = 4\times 10^{52}{\rm ergs~s^{-1}}$ in the
beginning of the afterglow phase. The resulting bulk Lorentz factor in
the shocked region is therefore 
\begin{equation}
\Gamma\approx 87L_{52.6}^{1/4}A_*^{-1/4}.
\end{equation}
The RS crosses the ejecta at the radius
\be
R_\times ({\rm wind}) \simeq 2.7\times 10^{15}{\rm cm}L_{52.6}^{1/2}
A_*^{-1/2}\Delta_{11.5}. 
\ee
At this radius, the rear of the MeV photon flow only leads the rear of
the ejecta by a distance of $R_\times({\rm wind}) /2\Gamma_0^2 \ll
\Delta$, so that {\em during most of the time when RS
crosses the ejecta, the shocked regions are still precipitated by the
MeV photons, so that the electrons and $e^\pm$ pairs in the shocks
lose energy mainly by scattering with the MeV photons}.  As a result,
the conventional synchrotron ultraviolet-optical emission of the
electrons in the RS region would be greatly suppressed. Let us define
a parameter $Y\equiv U_{\rm \gamma}/U_{\rm B}$, where
$U_{\gamma}\simeq L_{\gamma}/4\pi R^2\Gamma^2c$ is the MeV photon
energy density in the rest frame of the shocked region, and $U_{\rm
B}=B^2/8\pi\simeq (\Gamma_0/\Gamma)^2 \epsilon_B n_4 m_p c^2$ is the
comoving magnetic energy density in the same region, where $n_4=
(1-\eta) L /4\pi R^2 \Gamma_0^2 m_p c^3$ is the comoving density of
the un-shocked ejecta.  As usual, we introduce magnetic field and
electron equipartition parameters $\epsilon_{\rm B}\sim 0.001-0.1$ and
$\epsilon_{\rm e}\sim 0.1$ in the shocked region. The wide range of
$\epsilon_B$ is adopted with the consideration of broadband afterglow
fits (e.g. Panaitescu \& Kumar 2001) and the possibility that the
ejecta may be magnetized (e.g. Fan et al. 2002; Zhang et al. 2003). In
this work, we normalize our expression by taking $\epsilon_{\rm B}\sim
0.01$. After some simple algebra, we have
\be
Y\approx \eta /[(1-\eta )\epsilon_{\rm B}].
\label{Y}
\ee
For $\epsilon_{\rm B}\sim 0.001-0.1$ and $\eta\approx 0.2$, we have
$Y\sim 2.5-250\gg 1$. The cooling frequency $\nu_{\rm c}$ (see
Eq. (15) of DL01) would then be lowered by a factor $1/(1+Y)^2$.  For
$\epsilon_{\rm B,-2}\sim 0.1-1$, the flux of the soft
ultraviolet-optical photons is significantly lower than the one
calculated in DL01, so that the energy loss fraction for each proton
through pion production (i.e. $f_\pi$, defined as the ratio between
the comoving expansion time $t'_{\rm d}$ and the comoving proton
cooling time $t'_\pi$ due to photomeson cooling) should be also
lowered by roughly a factor of $1/(1+Y)$ (see equations [21] and [22]
in DL01). Consequently, the resulting detectable neutrino flux should
be somewhat lower.

It is worth pointing out that for some other processes accompanying
prompt $\gamma-$ray emission, such as the prediction of the
ultraviolet flash powered by neutron-rich internal shocks (Fan \& Wei
2004), the MeV photon flow also plays an imporatnt role on the cooling
of the accelerated electrons. By taking into account the overlapping
effect, Fan et al (2005b) have shown that with reasonable parameters,
the near IR, optical and UV emission flux from the neutron-rich
internal shocks can be as strong as 10mJy. Comparing with the standard
internal shock models, it also suffers weaker
synchrotron-self-absorption and seems to be able 
to account for prompt long-wavelength observation of GRB 041219a 
(e.g., Blake et al. 2005; Vestrand et al. 2005).  

\section{sub-GeV photon flash}\label{Sub-GeV}
Beloborodov (2005) suggested that electrons accelerated by the RS
would scatter the MeV photon flow and power a detectable GeV-TeV
flash. It is suggested that this mechanism may potentially acount for
the distinct high energy spectral component observed in GRB 941017
(Gonz\'{a}lez et al. 2003).

One uncertainty is that, if in the RS region $\epsilon_{\rm B}^r \geq
0.1$, as might have been the case for GRB990123 (Fan et al. 2002;
Zhang et al. 2003), 
the $Y$ parameter (equation [\ref{Y}]) is not large, and the IC effect
may not be strong. (Hereafter the superscripts $r$ and $f$
represent the RS and FS, respectively. If the parameters in both
regions are the same, no superscript is marked.)
Below we will show that by taking into account the
photon-shock interaction in the FS region, the IC process is generic
regardless of the magnetization of the ejecta. Also, the typical
energy of the emerged spectrum peaks in the sub-GeV range (see below).

At $R_{\times}({\rm wind})$, the energy contained in the RS region
can be written as
\begin{equation}
E_{\rm tot}^{\rm r}\approx \gamma_{34}\Gamma M_{\rm ej}c^2\approx
\Gamma_0 M_{\rm ej}c^2/2\approx E_{\rm tot}^{\rm f},
\label{E_totr}
\end{equation}
where we have taken the Lorentz factor of the shocked ejecta
relative to the unshocked ejecta as $\gamma_{34} \approx 
(\Gamma_0/\Gamma+\Gamma/\Gamma_0)/2\simeq \Gamma_0/2\Gamma \gg 1$
(which is valid for the wind RS model). Here
$M_{\rm ej}\simeq E/(\Gamma_0 c^2)$ is the rest mass of the
ejecta.
Equation (\ref{E_totr}) suggests that the energy
contained in the RS region is roughly equal to that in the FS region. 
As a result, if a significant part of the FS energy is emitted in the
sub-GeV range (we'll show below that it is the case), whether the
sub-GeV emission from the RS region is suppressed or not is
observationally unimportant. In view of this fact, below we adopt
the same physical parameters ($\epsilon_{\rm e}\sim 0.1$ and
$\epsilon_{\rm B}\sim 0.01$) for both the RS and the FS regions. 

The total thermal energies in the RS and FS regions are $E_{\rm th}^{\rm
r}\simeq (\gamma_{34}-1)E_{\rm tot}^{\rm r}/\gamma_{34}$ and $E_{\rm
th}^{\rm f}\simeq (\Gamma-1) E_{\rm tot}^{\rm f}/\Gamma$, respectively.
The ratio of them is $E_{\rm th}^{\rm r}/E_{\rm th}^{\rm f} \approx
(\gamma_{34}-1)/\gamma_{34}$. For 
typical parameters, $\gamma_{34}\sim 2$, we have $E_{\rm th}^{\rm
r}:E_{\rm th}^{\rm f}\approx 1:2$. This result suggests that for both
the high energy photon emission and the neutrino emission (as
discussed in \S4), the component from the FS region is the dominant
one. 

As usual, we assume that the shocked electrons distribute as $dN_{\rm
e}/d\gamma_{\rm e}\propto
\gamma_{\rm e}^{\rm -p}$ ($p\sim 2.2$) for $\gamma_{\rm m}<\gamma_{\rm
e}<\gamma_{\rm M}$, where $\gamma_{\rm M}\sim 10^8/B^{1/2}$ is the
maximum Lorentz factor of the electrons accelerated by
shocks, B is the magnetic field strength of the shock regions, which
can be calculated by 
\be
B\approx 1.86\times 10^3{\rm G}~\epsilon_{\rm
B,-2}^{1/2}A_*^{-1/4}L_{52.6}^{1/4}R_{15}^{-1}
\ee 

For the FS and RS, $\gamma_{\rm m}$ can be estimated by
\bea
\gamma_{\rm m}^{\rm f}&=&(\Gamma-1)\epsilon_{\rm e}(m_{\rm p}/m_{\rm
e})[(p-2)/(p-1)]\nonumber\\ 
&\approx &2600 \epsilon_{\rm e,-1}A_*^{-1/4}L_{52.6}^{1/4},
\nonumber\\
\gamma_{\rm m}^{\rm r}&=&(\gamma_{34}-1)\epsilon_{\rm e}(m_{\rm
p}/m_{\rm e})[(p-2)/(p-1)]\nonumber\\ 
&\approx & 23 \epsilon_{\rm e,-1}A_*^{1/4}L_{52.6}^{-1/4}
\Gamma_{0,2.5}.
\label{gamma_m}
\eea
Performing IC correction on equation (14) of DL01, we 
also get the cooling Lorentz factor $\gamma_{\rm c}\leq 1$,  
which means that the shocked
electrons all cool so rapidly that essentially all of them have
$\gamma_{\rm e}\sim 1$. 
In other words, all electrons are in the fast cooling phase.

The electrons in the shocked regions scatter with the MeV photons
and produce high energy photons. Given the observed prompt GRB peak
energy $\epsilon_{\rm \gamma,obs}^{\rm b}=300$keV (throughout the
paper, the subscript ``obs" denotes the parameters measured in the
observer frame), the electron energy above which the IC process is
suppressed by the Klein-Nishina limit is given by (see also
Beloborodov 2005)
\be
\gamma_{\rm e}^*={\Gamma m_{\rm e}c^2\over
(1+z)\epsilon^{\rm b}_{\rm \rm \gamma,obs}}\sim \Gamma.
\label{gamma_e*}
\ee
In the RS region, one has $\gamma_{\rm e}^* > \gamma_{\rm m}^{\rm r}$.
The peak energy of the upscattered photons can be then estimated as
\bea
\epsilon_{\rm m,obs}^{\rm r,IC} &\approx & {\gamma_{\rm m}^{\rm
r}}^2\epsilon^{\rm b}_{\rm \rm \gamma,obs}\nonumber\\ 
& \simeq & 0.16{\rm GeV}~\epsilon_{\rm e,-1}^2
A_*^{1/2}\Gamma_{0,2.5}^2L_{52.6}^{-1/2}({\epsilon_{\gamma,
\rm obs}^{\rm b}\over 300{\rm keV}}). 
\label{e_mrIC}
\eea
In the FS region, on the other hand, one has
$\gamma_{\rm m}^{\rm f}\gg \gamma_{\rm e}^*$, so that the
IC cooling by the MeV photons is suppressed significantly. The
fresh energetic electrons lose energy mainly by
synchrotron self-Compton (SSC) effects, the corresponding Compton
parameter can be estimated by 
$Y_{\rm SSC}\simeq \sqrt{\epsilon_{\rm e}/\epsilon_{\rm B}}\gg 1$.
The typical energy of SSC photons reads
\bea
 \epsilon_{\rm m,obs}^{\rm f,SSC}&\approx & h({\gamma_{\rm m}^{\rm
 f}})^4 e \Gamma B/[2\pi m_{\rm e}c(1+z)]\nonumber\\ &\approx &
 86{\rm GeV}~(1+z)^{-1}\epsilon_{\rm e,-1}^4\epsilon_{\rm
 B,-2}^{1/2}A_*^{-3/2}\nonumber\\ &&L_{52.6}^{3/2}R_{15}^{-1},
\label{e_mfSSC}
\eea
where $h$ is the Planck's constant.

Apparently, equations (\ref{e_mrIC}) and (\ref{e_mfSSC}) are quite
different. However, as shown below, photons more energetic than $\sim
1$ GeV will be absorbed by the MeV photons near the shock crossing
radius, and most energy will reprocessed and end up in the sub-GeV
range. This is valid for both 
the RS and FS. The pair production optical depth for photons with
energy $\sim 1$ GeV (which will be absorbed by the soft photons with
energy $\epsilon_{\rm a,obs}\simeq 2(\Gamma m_{\rm
e}c^2)^2/(1+z)^2{\rm GeV}\sim 1{\rm MeV}$) is roughly (e.g. Svensson
1987)
\bea
\tau_{\gamma\gamma}(1{\rm GeV})&\approx &   
{11\sigma_{\rm T}N_{>\epsilon_{\rm a,obs}}\over 720\pi R^2}\nonumber\\
&\approx & 20E_{\gamma,53}({1+z\over 2})L_{52.6}^{-1/2}A_*^{1/2}
R_{15}^{-2}, 
\label{tau}
\eea
where $N_{>\epsilon_{\rm a,obs}}$ is the total number of photons
satisying $\epsilon_{\rm obs}>\epsilon_{\rm a,obs}$, which can be
estimated by $N_{>\epsilon_{\rm a,obs}}\approx
0.2E_{\gamma}/[(1+z)\epsilon_{\rm a,obs}]$ for 
$\epsilon_{\rm a,obs}>\epsilon_{\rm \gamma,obs}^{\rm b}$, where the coefficient
$0.2$ is a rough spectrum correction factor on the photon number
calculation derived from the integration of the GRB spectrum. In this
paper, we take a typical broken power law GRB spectrum for
MeV photons, i.e., $n(\epsilon_{\gamma})\propto \epsilon_\gamma^{-1}$
for $\epsilon_{\gamma}^{\rm b}/50<\epsilon_{\gamma}<
\epsilon_{\gamma}^{\rm b}$, and
$n(\epsilon_{\gamma})\propto \epsilon_\gamma^{-2}$ for
$\epsilon_{\gamma}^{\rm b}<\epsilon_{\gamma}<50\epsilon_{\gamma}^{\rm b}$. 

With typical parameters, we have $\tau_{\gamma\gamma}({\rm
1GeV})\simeq 2$ at $R_\times ({\rm wind})$. This means that photons
more energetic than 1GeV are absored by the MeV
photons and produce $e^\pm$ pairs. The random Lorentz factor of
these secondary pairs can be estimated as $\gamma_{\pm}\simeq
(1+z)\epsilon_{\rm \gamma, 
obs}/2\Gamma m_{\rm e}c^2=11(1+z)(\epsilon_{\rm \gamma, obs}/1{\rm
GeV})$. For $\epsilon_{\rm \gamma, obs}=\epsilon_{\rm \gamma, obs}^{\rm
f, SSC}\sim 32(1+z)^{-1}{\rm GeV}$, we have $\gamma_{\pm}^{\rm
f,SSC}\simeq 350$. These pairs will scatter both the MeV photons and
the X-ray photons produced in the forward shock region.
Since $\gamma_{\pm}^{\rm
f,SSC}$ is significantly larger than $\gamma_{\rm
e}^*$, so that the Klein-Nishina correction is still important for the
pair$-$MeV photon IC scattering. The 
cross section can be approximated as $\sigma\approx A(x)\sigma_{\rm
T}$, where $x\simeq \gamma_\pm^{\rm f,SSC}/\gamma_{\rm e}^*\sim
\gamma_\pm^{\rm f,SSC}/\Gamma\approx 4$, and $A(x)\approx {3\over 8x}({\rm
ln}2x+{1\over 2})\approx 0.2$\footnote{In the Klein-Nishina regime, for one electron,
 the energy loss 
in one scattering is $\sim \gamma_{\rm e}m_{\rm e}c^2$, the energy loss rate 
is $P_{\rm IC}\sim \gamma_{\rm e}m_{\rm e}c^2 A(x)\sigma_{\rm T} n_{\gamma} c$, where 
$n_\gamma\sim U_\gamma/(\epsilon_\gamma^{\rm b}/\Gamma)$ is the number density of $\gamma-$ray photons.
 The synchrotron radiation power satisfies $P_{\rm syn}=(4/3)\sigma_{\rm T}
\gamma_{\rm e}^2 \beta_{\rm e}^2U_{\rm B}c$. The Compoton parameter can be estimated by $Y_{_{\rm KN}}
=P_{\rm IC}/P_{\rm syn}\sim A(x)U_\gamma/(xU_{\rm B})$, where we have taken
 $\gamma_{\rm e}=\gamma_\pm^{\rm f,SSC}$. Since $U_\gamma/U_{\rm B}$ is just the 
Compton parameter $Y$ in the Thomson regime, we have $Y_{_{\rm KN}}\sim A(x)Y/x$ for $x\gg 1$, i.e., 
in the extreme Klein-Nishina regime the IC cooling is 
 very inefficient.}.
The energy density of the soft X-rays (FS synchrotron emission) is
about $g=(1-\eta)\epsilon_{\rm e}/[ 
\eta(1+Y_{\rm SSC})]\sim 0.1$ times of that
of the MeV photons, so the fraction of energy of the secondary pairs
that goes to scattering with the soft X-ray FS emission 
(with a typical frequency defined by eq.
[\ref{nu_mf}]) 
can be estimated as $[4g/A(4)]/[1+4g/A(4)]\sim 2/3$. The typical energy
of the upscattered soft X-rays can be estimated as (at
$R_\times ({\rm wind})$) 
\bea
\epsilon_{\rm m,obs}^{\rm f,IC}&\simeq & (\gamma_\pm^{\rm
f,SSC})^2\epsilon_{\rm m,obs}^{\rm f} 
 \simeq  0.29 {\rm GeV} ({\gamma_\pm^{\rm f,SSC}\over 350})^2\nonumber\\
&&({1+z\over 2})^{-1}\epsilon_{\rm e,-1}^2\epsilon_{\rm
B,-2}^{1/2}A_*^{-1}L_{52.6}. 
\eea
The pairs also scatter the initial MeV photons. However, the
upscattered photons are still too energetic to escape. A pair photon
cascade likely develops until the photon energy is low enough to be
transparent. For example, in the extreme
Klein-Nishina limit, the typical energy of the upscattered MeV
photons can be estimated by $(1+z)\epsilon_{\rm obs}^{\rm IC}\sim
\Gamma \gamma_\pm^{\rm f,SSC} m_{\rm e}c^2 \sim 15.6 {\rm
GeV}~L_{52.6}^{1/4}A_*^{-1/4}(\gamma_\pm^{\rm
f,SSC}/350)$. After interacting with the MeV photons again,
tertiary pairs will be produced with a typical Lorentz factor of $\sim
180$. They then re-scatter various photon fields again. 
The detailed investigation of this cascade process has to be modeled
numerically, and is beyond the scope of this paper. Nonetheless, a
generic picture is that at least 1/3 of the total energy of the
reprocessed photons is radiated in the sub-GeV range, making these
GRBs interesting targets for GLAST observations .

For typical parameters taken in this work ($z=1$ and $Q_{\rm x}=1$),
the fluence of the sub-GeV flash from the very early FS can be
estimated as
\bea
S&\sim &\frac{2}{3}\epsilon_{\rm e}\Gamma (\Gamma-1)m_{\rm p}c^2(1+z)
\int_0^{R_\times ({\rm 
wind})} R^2 ndR/ D_{\rm L}^2\nonumber\\ &\approx & 2\times 10^{-6}~{\rm ergs~cm^{-2}},
\label{S}
\eea
where $D_{\rm L}$ is the luminosity distance, which $\approx 2.2\times
10^{28}$cm for $z=1$. The fluence for the RS component may be
comparable or less than this value.
For EGRET, the estimated fluence threshold in
the low integration time ($t_{\rm int}$) regime ($t_{\rm
int}<1.7\times 10^3 {\rm s}$) is $\sim 2.1\times 10^{-6}~{\rm
ergs~cm^{-2}}$ for a typical photon energy $\sim 400$ MeV (e.g. Zhang
\& M\'{e}sz\'{a}ros 2001). Our predicted fluence $S\sim 
10^{-6}~{\rm ergs~cm^{-2}}$ at sub-GeV energy range is below the
EGRET detection threshold, which explains the none detection of such
flashes in the EGRET era. The fluence threshold of GLAST is roughly $\sim
4\times 10^{-7}~{\rm ergs~cm^{-2}}$ for $t_{\rm int}<10^{5}{\rm s}$
(e.g. Zhang \& M\'{e}sz\'{a}ros 2001). We therefore expect that many
sub-GeV photon flashes can be detected in the GLAST era.

\section{Neutrino emission}
\label{neutrinos} 

\subsection{Neutrinos from photomeson interaction with the
synchrotron photons in RS and FS}\label{10^15}

The protons accelerated from both the FS and the RS would produce high
energy neutrinos via photomeson interaction, mainly through
$\Delta$-resonance. In the RS region, DL01 have found that the typical
neutrino energy is $3\times 10^{15}-3\times 10^{17}$ eV.

We now estimate the neutrino emission from the FS region.
As mentioned in \S{\ref{Sub-GeV}} (see the paragraph below equation
[\ref{e_mrIC}]), the fresh electrons accelerated by the FS lose
energy mainly by SSC cooling rather than by IC cooling. As a result,
the flux above $\nu_{\rm m}^{\rm f}$ should be lowered by a factor
$1/(1+Y_{\rm SSC})$. The typical synchrotron radiation frequency
($\nu_{\rm m}^{\rm f}$) of the FS reads
\bea
\nu_{\rm m}^{\rm f}\approx 3.1\times 10^{18}{\rm Hz}~\epsilon_{\rm
e,-1}^2\epsilon_{\rm B,-2}^{1/2} 
L_{52.6}A_*^{-1}R_{15}^{-1},
\label{nu_mf}
\eea
 
The protons that interact with the $h\nu_{\rm m}^{\rm f}$ photons at
the $\Delta-$resonance have an observed energy of
\bea
\epsilon_{\rm p,obs}^{\rm b}(1)&\approx & 0.3{\rm
GeV}^2\Gamma^2/[(1+z)h\nu_{m}^{\rm f}]\nonumber\\ 
 &\approx & 1.8\times 10^{17}{\rm eV}~(1+z)^{-1}\epsilon_{\rm
 e,-1}^{-2}\epsilon_{\rm B,-2}^{-1/2}\nonumber\\ &&L_{52.6}^{-1/2}
 A_*^{1/2}R_{15}.
\eea
 The photomeson interactions include (1) production of $\pi$ mesons,
 $p\gamma\rightarrow p+\pi^0$ and $p\gamma \rightarrow n+\pi^+$, and
 (2) decay of $\pi$ mesons, $\pi^0\rightarrow 2\gamma$ and
 $\pi^+\rightarrow \mu^+ +\nu_{\mu}\rightarrow e^+ +\nu_{\rm e}+\bar
 \nu_{\rm \mu}+\nu_{\mu}$.  When protons interact with the FS 
 photons with typical energy $h\nu_{\rm m}^{\rm f}$, these
 processes produce neutrinos with energy $\epsilon^{\rm b}_{\rm
 \nu,obs}(1) \sim 0.05\epsilon^{\rm b}_{\rm p,obs}(1)$ (e.g. Waxman \&
 Bahcall 1997), i.e.
\bea
\epsilon^{\rm b}_{\nu,\rm obs}(1)&\simeq & 9\times 10^{15}{\rm
eV}~(1+z)^{-1}\epsilon_{\rm B,-2}^{-1/2}\epsilon_{\rm
e,-1}^{-2}\nonumber\\ 
&&L_{52.6}^{-1/2}A_*^{1/2}R_{15}.
\eea
In principle, $\gamma-$rays of similar energies are produced by
$\pi^0$ decay. But as shown in \S\ref{Sub-GeV}, the MeV photon flow is
optically thick for these high energy photons, and a $e^\pm$ pair
cascade develops. The bulk $\gamma-$rays emerge at much lower energies
$\sim {\rm GeV}$, which adds to the sub-GeV emission discussed in
\S\ref{Sub-GeV}. 

Similar to Halzen \& Hooper (2002) and Guetta et al.(2004), at
$\epsilon_{\rm \nu,obs}^{\rm b}(1)$, the fraction of energy loss of
protons through photomeson interaction with the $h\nu_{\rm m}^{\rm f}$
photons, $f_\pi(1)$, can be paramterized as (in the calculation, we
have taken $Y_{\rm SSC}\approx 3.2\epsilon_{\rm
e,-1}^{1/2}\epsilon_{\rm B,-2}^{-1/2}$)
\bea
f_\pi(1) &\sim & \sigma_\Delta~ n_\gamma(1) \Delta R'<x_{\rm
p\rightarrow \pi}>
\nonumber\\
&\approx & {\epsilon_{\rm e}E \over 2 (1+Y_{\rm SSC})
(h\nu_{\rm m}^{\rm f})4\pi R R_\times}\sigma_\Delta <x_{\rm
p\rightarrow \pi}>\nonumber\\ &\approx & 0.66\epsilon_{\rm
e,-1}^{-3/2}L_{52.6}^{-1/2}A_*^{3/2},
\label{f_pi1} 
\eea
where $\sigma_\Delta \approx 5\times 10^{-28}{\rm cm^2}$ is the cross
section of the $\Delta-$resonance, $<x_{\rm p\rightarrow \pi}>\simeq 0.2$
is the average fraction of the energy transfered from the initial proton
to the produced pion, $\Delta R'$ is the width of the shock, and
$n_\gamma(1)\approx (R/R_\times) [\epsilon_{\rm e}E/ (2 h\nu_{\rm
m}^{\rm f})]/[4\pi(1+Y_{\rm SSC}) R^2\Delta R']$ (for 
$R\leq R_\times$) is the photon number density, $E\sim 4\times
10^{53}{\rm ergs}$  
is the isotropic energy of the ejecta at the beginning of the afterglow, 
as defined in \S{\ref{overlapping}}. It is interesting to
see that $f_\pi(1)$ is independent on $t$ since $\nu_{\rm m}^{\rm
f}\propto R^{-1}$.

Below we estimate the maximum proton energy ($\epsilon_{\rm p}^{\rm
M}$) accelerated by the shocks. In general, $\epsilon_{\rm p}^{\rm
M}={\rm min}[\epsilon_{\rm p}^{\rm M}(1), \epsilon_{\rm p}^{\rm M}(2),
\epsilon_{\rm p}^{\rm M}(3)]$ satisfies three constraints. (1) The
comoving shock acceleration time $t'_{\rm a}\sim \epsilon_{\rm
p}/\Gamma e B c$ should be smaller than the comoving wind expansion
time $t'_{\rm d}\sim R/\Gamma c$, which yields $\epsilon_{\rm p}^{\rm
M}(1)\sim e B R$. The numerical value reads
\be 
\epsilon_{\rm p}^{\rm M}(1)\simeq 5\times 10^{20}{\rm
eV}~\epsilon_{\rm B,-2}^{1/2}L_{52.6}^{1/4}A_*^{-1/4}.  
\label{eps_p1}
\ee
In this work, the superscript $'$  represents the parameter measured in
the comoving frame.

(2) The comoving proton synchrotron cooling timescale $t'_{\rm
cool}=[6\pi m_{\rm p}^4c^3/\sigma_{\rm T}m_{\rm e}^2]\Gamma
\epsilon_{\rm p}^{-1} B^{-2}$ should be longer than the comoving
acceleration timescale $t'_{\rm a}$ (e.g., Li et al. 2002), which
results in
\bea
\epsilon_{\rm p}^{\rm M}(2)\simeq  4\times 10^{20}{\rm eV}~
\epsilon_{\rm B,-2}^{-1/4}L_{52.6}^{1/8}
A_*^{-1/8}R_{15}^{1/2}.
\label{eps_p2}
\eea
(3) The comoving proton cooling timescale due to photomeson
interaction $t'_\pi$ should also be longer than the comoving
acceleration timescale $t'_{\rm a}$. $t'_\pi$ can be derived from
$\bar{f}_\pi \simeq t'_{\rm d}/t'_\pi$ (e.g. Waxman \& Bachall 1997), 
i.e., $t'_\pi\simeq (R/\Gamma c){\bar{f}_\pi}^{-1}$, where
$\bar{f}_{\pi}=f_\pi(1)(\epsilon_{\rm p}/\epsilon_{\rm p}^{\rm
b})^{1/2}$ for $\epsilon_{\rm p}>\epsilon_{\rm p}^{\rm b}(1)$.
Requiring $t'_\pi=t'_{\rm a}$ yields $\epsilon_{\rm p}/(\Gamma e B
c)=R/[\bar{f}_\pi \Gamma c]$, which can be simplified as $\epsilon_{\rm
p}^{\rm M}(3)\simeq {\bar{f}_\pi}^{-1} \epsilon_{\rm p}^{\rm
M}(1)\simeq f_\pi(1)[\epsilon_{\rm p}^{\rm M}(3)/\epsilon_{\rm p}^{\rm
b}]^{-1/2}\epsilon_{\rm p}^{\rm M}(1)$. We then have
\bea
\epsilon_{\rm p}^{\rm M}(3)&\approx & {f_\pi(1)}^{-2/3}[\epsilon_{\rm
p}^{\rm M}(1)]^{2/3}(\epsilon_{\rm p}^{\rm b})^{1/3}\nonumber\\ 
&\simeq & 4\times 10^{19}{\rm eV}~
\epsilon_{\rm e,-1}^{1/3}\epsilon_{\rm B,-2}^{1/6}L_{52.6}^{1/3}
A_*^{-1}R_{15}^{1/3}, 
\label{eps_p3}
\eea 
Since $\epsilon_{\rm p,obs}^{\rm b} \ll \epsilon_{\rm p}^{\rm
M}=\epsilon_{\rm p}^{\rm M}(3)\sim 4\times 10^{19}$eV, the predicted
flux of neutrino emission is insensitive to the actual value of
$\epsilon_{\rm p}^{\rm M}$.

After the pions (muons) are generated, the high energy pions (muons) may
lose energy via synchrotron emission before decaying, thus reducing
the energy of the decay neutrinos (e.g. Guetta et al. 2004). For
pions, this effect should be taken into account when the pion lifetime
$\tau'_\pi\approx 2.6\times 10^{-8}\epsilon'_\pi/(m_\pi c^2){\rm s}$
is comparable to the synchrotron loss time $t'_{\rm \pi,syn}=6\pi
m_\pi^4c^3/(\sigma_{\rm T}m_{\rm e}^2\epsilon'_{\rm \pi}B^2)$. We
can then define a critical pion energy ${\epsilon'_\pi}^{\rm c}$ by 
requiring $\tau'_\pi=t'_{\rm \pi,syn}$, which gives
\be
{\epsilon'_{\rm \pi}}^{\rm c}\approx 1.1\times 10^{17}{\rm eV}~
\epsilon_{\rm B,-2}^{-1/2}L_{52.6}^{-1/4}A_*^{1/4}R_{15},
\ee
above which the neutrino emission is suppressed significantly.
Correspondingly, we have a critical observed energy for $\nu_{\rm \mu}$
(since ${\epsilon'}_{\rm \pi}^{~\rm c}\approx
4{\epsilon'}_{\nu_\mu}^{\rm ~c}$),
\be
\epsilon_{\rm \nu_{\rm \mu},obs}^{\rm c}\approx {\Gamma
{\epsilon'_{\rm \pi}}^{\rm c}\over 4(1+z)}\approx 1.2\times
10^{18}{\rm eV}~({1+z\over 2})^{-1}\epsilon_{\rm B,-2}^{-1/2}R_{15}, 
\label{e_nu}
\ee
above which the slope of the $\nu_{\mu}$ spectrum steepens by 2 since
$t'_{\rm \pi,syn}/\tau'_{\rm \pi}\propto {\epsilon'_\pi}^{-2}$.

Muons have a lifetime $\sim 100$ times longer than that of pions, so
that the energy cutoff of $\bar{\nu}_{\rm \mu}$ and $\nu_{\rm e}$ will
therefore be 10 times smaller than $\epsilon_{\rm \nu_{\rm
\mu},obs}^{\rm c}$ (e.g. Guetta et al. 2004). With equation
(\ref{e_nu}) we have
\be
\epsilon_{(\rm \bar{\nu}_{\rm \mu},\nu_{\rm e}), obs}^{\rm c}\approx
{\epsilon_{\rm \nu_{\rm \mu},obs}^{\rm c}\over 10}\approx 1.2\times
10^{17}{\rm eV}~({1+z\over 2})^{-1}\epsilon_{\rm B,-2}^{-1/2}R_{15}, 
\label{e_barnu}
\ee

Equation (\ref{f_pi1}) is valid only for protons at the break energy
$\epsilon^{\rm b}_{\rm p, obs}(1)$. Considering emission from all the
protons, the resulting neutrino spectrum would trace the photon
spectrum. Generally one has
\bea
&& f_\pi \approx
\left\{ 
  \begin{array}{@{\,}lc} f_\pi(1)[{\epsilon_{\rm p,obs}\over
  \epsilon_{\rm p,obs}^{\rm b}(1)}]^{1.1}, & \epsilon_{\rm
  p,obs}<\epsilon_{\rm p,obs}^{\rm b}(1),\\

   f_\pi(1)[{\epsilon_{\rm p,obs}\over \epsilon_{\rm p,obs}^{\rm
   b}(1)}]^{0.5}, & \epsilon_{\rm p,obs}^{\rm b}(1)<\epsilon_{\rm
   p,obs}<\epsilon_{\rm p,obs}^{\rm 0},\\

  1, & \epsilon_{\rm p,obs}^{\rm 0}<\epsilon_{\rm p,obs}<
  \epsilon_{\rm p,obs}^{\rm M}, \end{array} \right.
\label{Spectrum1}
\eea
where $\epsilon_{\rm p,obs}^0$ is defined as $(\epsilon_{\rm
p,obs}^0/\epsilon_{\rm p,obs}^{\rm b})^{0.5}=1/f_\pi(1)$;
$\epsilon_{\rm p,obs}^{\rm M}=\epsilon_{\rm p}^{\rm M}/(1+z)$.
The neutrino spectrum finally reads
\bea
&& {\epsilon_{\rm \nu,obs}^2 d^2N_{\rm \nu}\over d\epsilon_{\rm
\nu,obs} dt} \simeq {1\over 8} {(1+z)L \over 4\pi D_{\rm
L}^2}\nonumber\\ &&
\left\{ 
  \begin{array}{@{\,}lc} f_\pi(1)[{\epsilon_{\rm \nu,obs}\over
  \epsilon_{\rm \nu,obs}^{\rm b}(1)}]^{1.1}, & \epsilon_{\rm
  \nu,obs}<\epsilon_{\rm \nu,obs}^{\rm b}(1),\\

   f_\pi(1)[{\epsilon_{\rm \nu,obs}\over \epsilon_{\rm \nu,obs}^{\rm
   b}(1)}]^{0.5}, & \epsilon_{\rm \nu,obs}^{\rm b}(1)<\epsilon_{\rm
   \nu,obs}<\epsilon_{\rm \nu,obs}^{\rm 0},\\

  1, & \epsilon_{\rm \nu,obs}^{\rm 0}<\epsilon_{\rm \nu,obs}<
  \epsilon_{\rm \nu,obs}^{\rm c},\\

({\epsilon_{\rm \nu,obs}\over \epsilon_{\rm \nu,obs}^{\rm c}})^{-2}, &
\epsilon_{\rm \nu,obs}^{\rm c}<\epsilon_{\rm \nu,obs}<\epsilon_{\rm
\nu,obs}^{\rm M}.  \end{array} \right.
\label{Spectrum1}
\eea
where the factor 1/8 takes into account the fact that charged and
neutral pions are produced with roughly equal probabilities, and that
each neutrino carries $\sim 1/4$ of the pion energy, i.e. $\epsilon_{\rm
\nu,obs}^0\approx 
0.05\epsilon_{\rm p,obs}^0$. In equation (\ref{Spectrum1}),
$\epsilon_{\rm \nu,obs}^{\rm c}<\epsilon_{\rm \nu,obs}^{\rm M}$ is
assumed. In case of 
$\epsilon_{\rm \nu,obs}^{\rm c}>\epsilon_{\rm \nu,obs}^{\rm M}$, 
the neutrino flux is 0 for $\epsilon_{\rm
\nu,obs}>\epsilon_{\rm \nu,obs}^{\rm M}$.

As in many publications, we assume that the protons in the FS
distribute as $dn/d\gamma_{\rm p}\propto \gamma_{\rm p}^{-2}$ for
$\gamma_{\rm p,m}<\gamma_{\rm p}<\gamma_{\rm p,M}$, where $\gamma_{\rm
p,m}=(1-\epsilon_{\rm e})(\Gamma-1)/{\rm ln}(\gamma_{\rm
p,M}/\gamma_{\rm p,m})$ and $\gamma_{\rm p,M}={\epsilon}_{\rm p}^{\rm
~ M}/\Gamma m_{\rm p}c^2$.  For a neutrino detector with a surface
area of $S_{\rm det} \sim {\rm km^2}$, the detectable events for one
burst can be estimated by (taking $R=0.5R_\times ({\rm wind})$)
\bea
 N_{\rm events}(1)\approx {1\over 8} {(1-\epsilon_{\rm e})E \over
\epsilon_{\rm \nu,obs}^{\rm b}(1)}{1\over 4\pi D_{\rm 
L}^2}S_{\rm det} P_{\nu \rightarrow \mu}\nonumber\\ 
{{\rm
ln[\epsilon_{\rm \nu,obs}^{\rm 0}/\epsilon_{\rm \nu,obs}^{\rm
b}(1)]f_\pi(1)+{\rm ln}(\epsilon_{\rm \nu,obs}^{\rm c}/\epsilon_{\rm
\nu,obs}^{\rm 0})}\over {\rm ln}(\gamma_{\rm p, M}/\gamma_{\rm
p,m})}\nonumber\\
\sim 0.002E_{53.6}D_{\rm L,28.34}^{-2},
\label{N_events1}
\eea 
where $P_{\nu \rightarrow \mu}\simeq 6\times 10^{-4}(\epsilon_{\rm
\nu,obs}/3\times 10^{15}{\rm eV})^{0.5}$ is the probability that a
neutrino produces a detectable high energy muon for $\epsilon_{\rm
\nu,obs}>10^3{\rm TeV}$ (Gaisser et al. 1995). The factor $1/8$ takes
into account the fact that there are three neutrinos produced with
each carrying about $1/8$ of the energy loss of the proton, and that
due to neutrino flavor oscillation, only 1/3 of the neutrinos are the
detectable muon neutrinos.
For a nearby GRB with $z\sim 0.1$ ($D_{\rm L}\sim 1.4\times 10^{27}$
cm) and $E\sim 4\times 10^{53}$ergs, the resulting
detectable events is $\sim 0.4$. 
For such a nearby energetic GRB, the prompt $\gamma-$ray energy
fluence would be ${\cal F}_\gamma = (1+z)E_\gamma/(4\pi D_{\rm
L}^2)\sim 10^{-3} {\rm ergs~cm^{-2}}E_{\gamma,53}D_{\rm
L,27.15}^{-2}$. This means that only for those very bright  
GRBs can the predicted neutrinos be detected. Similar conclusions 
have been also drawn in Dermer \& Atoyan (2003) and Guetta et
al. (2004). We note that low redshift long GRBs are rare.
So far we have detected at least three of them, but none of them is
energetic enough for the above purpose.

Another method to estimate the detectability of the neutrino
early afterglow emission is to assume that GRBs are the dominant
sources of high energy cosmic rays (from the internal or external
shocks, Waxman 1995; Vietri 1995). This leads to a detectability about 
an order of magnitude higher  (see e.g. DL01) than the value presented
above. One possible reason for the discrepancy is that this second
method is based on the assumption that ultrahigh energy cosmic rays
are mostly produced by GRBs, which might not be the case.

The atmospheric neutrino background flux is $\phi_{\rm \nu,bkg}\sim
10^{-12}(\epsilon_{\rm \nu,obs}/10^{14}{\rm eV})^{-5/2}{\rm
cm^{-2}~s^{-1}~{\rm sr}^{-1}}$. The typical angular resolution of the
planned neutrino telescopes at TeV energies is about 1 degree.
For a typical GRB with $T_{90}\sim 20 {\rm s}$, $E\sim 4\times
10^{53}$ergs and $z=1$, at $\epsilon_{\rm \nu,obs}\sim 5\times
10^{15}{\rm eV}$, the background neutrino number in this time span is
$\sim 5\times 10^{-9}$, which is $\ll 0.002$ (eq.[\ref{N_events1}]).  
If a $\sim 5\times 10^{15}$eV neutrino well correlated with one GRB
both in time and coordinates has been detected, the chance probability
of being an atomospheric neutrino is only $\sim 3\times 10^{-5}$. 
Therefore, these GRB neutrinos can be detected above the background
(see also M\'{e}sz\'{a}ros \& Waxman 2001).

\subsection{Neutrinos from photomeson interaction with the
MeV photon flow: $\sim 10^{14}$eV}\label{10^14}

Because of the significant overlapping of the MeV photon flow and the
shocked regions, high energy protons with energy larger than $2\times
10^{15}{\rm eV}$ accelerated from both the RS and the FS would
interact with the MeV flow effectively. We now study the neutrino
emission process in this case. 

In the rest frame of the shocked regions, $\gamma-$ray photons with
the observed typical energy $\epsilon^{\rm b}_{\gamma,
\rm obs}\approx 300{\rm keV}$ have an energy
$\epsilon^{\rm b}_{\gamma}\approx (1+z)\epsilon_{\gamma,\rm obs}^{\rm
b}/\Gamma$. The protons that interact with the photons with energy
$\epsilon^{\rm b}_{\gamma}$ at the $\Delta$-resonance have
an observed energy of
\bea
&\epsilon_{\rm p,obs}^{\rm b}(2)\approx 0.3{\rm
GeV}^2\Gamma^2/(1+z)^2\epsilon^{\rm b}_{\gamma, \rm
obs}\nonumber\\&=2\times 10^{15}{\rm eV}~L_{52.6}^{1/2}A_*^{-{1/2}}
({1+z\over 2})^{-2}({\epsilon_{\gamma, \rm obs}^{\rm b}\over 300{\rm
keV}})^{-1}, 
\eea  
This energy is well below the observed maximum proton energy
$\epsilon_{\rm p}^{\rm M}$ (equations
[\ref{eps_p1}]-[\ref{eps_p3}]). The generated
neutrinos have a typical energy
\bea
\epsilon^{\rm b}_{\nu,\rm obs}(2)\sim 10^{14}{\rm eV}~L_{52.6}^{1\over
2}A_*^{-{1\over 2}}({1+z\over 2})^{-2} ({\epsilon^{\rm b}_{\gamma,\rm
obs}\over {\rm 300keV}})^{-1}.
\eea

Similar to \S{\ref{10^15}}, at $\epsilon_{\rm p,obs}^{\rm b}(2)$, the
fraction of the energy loss of protons through photomeson interaction
with the MeV photons can be parameterized as
\bea
f_{\pi}(2)\approx 0.18 L_{\gamma,52}L_{52.6}^{-1/2}
A_*^{1/2}\left[{(1+z)\epsilon^{\rm b}_{\gamma,\rm
obs}\over {\rm 600keV}}\right]^{-1}R_{15}^{-1}.
\eea
The MeV flow photons has a broken power law spectrum, i.e.,
$n(\epsilon_\gamma)\propto \epsilon_\gamma^{-1}$ below $\epsilon_{\rm
\gamma,obs}$ and $n(\epsilon_\gamma)\propto \epsilon_\gamma^{-2}$
above $\epsilon_{\rm \gamma,obs}^{\rm b}$. So the general form of the
fraction of the energy 
loss of protons through photomeson interaction with the MeV photons is
\bea
f_{\pi}\approx f_\pi(2)\left\{ \begin{array}{@{\,}lc} 1, &
\epsilon_{\rm p,obs}>\epsilon_{\rm p,obs}^{\rm b}(2),\\ {\epsilon_{\rm
p,obs}\over \epsilon_{\rm p,obs}^{\rm b}(2)}~, & \epsilon_{\rm p,obs}
\leq \epsilon_{\rm p,obs}^{\rm b}(2)~.  \end{array} \right.
\label{f_pi2}
\eea
As a result, the differential neutrino spectrum is also a broken power
law, i.e.  $n(\epsilon_{\nu})\propto \epsilon_\nu^{-1}$ below
$\epsilon_{\nu,\rm obs}^{\rm b}(2)$, and $n(\epsilon_{\nu})\propto
\epsilon_\nu^{-2}$ above $\epsilon_{\nu,\rm obs}^{\rm b}(2)$ (see also
Waxman \& Bachall 1997).  However, for $\epsilon_{\rm
\nu,obs}>\epsilon_{\rm \nu,obs}^0$, the energy of the corresponding
protons has been lost mainly by interacting with the keV soft photons
from the FS region (see \S{\ref{10^15}} for detail).  So the neutrino
spectrum can not extend to $\epsilon_{\rm \nu,obs}>\epsilon_{\rm
\nu,obs}^0$.

Similar to equation (\ref{N_events1}), we now estimate the
detectability of such $10^{14}$eV neutrinos in the early afterglow
phase. For one typical GRB, taking $R=0.5R_\times$(wind), one has
\bea
N_{\rm events}(2)\approx {1\over 8} {(1-\epsilon_{\rm e})E\over
\epsilon_{\rm \nu,obs}^{\rm b}(2)}{1\over 4\pi D_{\rm 
L}^2}S_{\rm det} P_{\nu \rightarrow \mu}\nonumber\\ {{\rm
ln}[\epsilon_{\rm \nu,obs}^{\rm 0}/\epsilon_{\rm \nu,obs}^{\rm
b}(2)]f_\pi(2)\over {\rm ln}(\gamma_{\rm p, M}/\gamma_{\rm p,m})}\sim
0.001E_{53.6}D_{\rm L,28.34}^{-2},
\label{N_events2}
\eea 
where $P_{\nu \rightarrow \mu}\simeq 1.7\times 10^{-6}(\epsilon_{\nu,
\rm obs}/1{\rm TeV})^{0.8}$ (e.g., Halzen
\& Hooper 2002) for $1{\rm TeV}<\epsilon_{\nu, \rm obs}<10^3{\rm TeV}$.
The resulting detectability is comparable with that of $5\times
10^{15}$eV (see equation [\ref{N_events1}]). These neutrinos can be
only marginally detectable above the atmospheric neutrino background.
For a typical GRB with $T_{90}\sim 20{\rm s}$, $E\sim 4\times
10^{53}$ergs  and $z=1$, the $10^{14}$eV neutrino background event
number is $\sim 9\times 10^{-5}$ in the time span. It is $< 
0.001$ (eq.[\ref{N_events2}]), but not significantly. For a detected
$\sim 10^{14}$eV neutrino well correlated with one GRB, the chance
probability of being an atomospheric neutrino is $\sim 10\%$. 

The total detectable events for one typical GRB born in the wind
environment ($z=1$) is $\sim 0.003$, this rate can be also enhanced
significantly when fluctuations in distance, energy are considered
(e.g., Haltzen \& Hooper 2002). Therefore, nearby GRBs such as GRB
030329 are of great interest (e.g., Razzaque et al. 2004). Thanks to
the correlation of the 
neutrinos with GRBs, both in time and coordinates, these GRB neutrinos
are likely to be detected with the planned ${\rm km^3}$ high energy
neutrino detectors such as ICECUBE, ANTARES, and NESTOR.

One caveat is that the $\sim 10^{14}$eV neutrinos would also overlap
with the $\sim 5\times 10^{14}$eV neutrinos generated from the
internal shocks (e.g., Waxman \& Bachall 1997), so that it might not
be easy to distinguish both species. We notice, however, that if GRB
ejecta are strongly magnetized so that the prompt emission is not
powered by internal shocks, the $10^{14}$ eV neutrinos discussed here
are still produced. 

\section{Discussion and conclusion}\label{DisCon}
Massive stars are likely the progenitor of long, soft GRBs.  For an
ultra-relativistic fireball emerging from a massive star and
expanding into a dense stellar wind medium, a strong relativistic RS
develops so early that both the RS and FS regions overlap in time
and space with the prompt MeV photon flow.  This overlapping leads to
modification of the early RS and FS emission significantly, since the
dominant cooling process for electrons is likely the IC with the MeV
photons (especially for the RS region,
Beloborodov 2005; see also Fan et al. 2005a for a more detailed
calculation on the modification of very early R-band lightcurve).

Due to the tight overlapping of the MeV photon flow and the shocked
regions (both FS and RS), the optical depth of the multi-GeV
photons is larger than unity. These high energy photons (SSC and/or IC
emission components from the FS and RS regions) are absorbed by the MeV
photon flow and therefore generate relativistic $e^\pm$ pairs. These
$e^\pm$ pairs re-scatter the soft X-rays in the FS region and the prompt
$\gamma-$rays and power detectable high energy emission. A significant
portion of the reprocessed photon energy is distributed in the sub-GeV
band, making wind-interaction GRBs interesting targets for the GLAST.
In the wind model, the total thermal
energy contained in the FS region is larger or at least comparable
with that contained in RS region. The above effect is therefore
generic, regardless of the possible magnetization of the ejcta
 (e.g. Fan et al 2002; Zhang et al 2003, in which case
the IC emission as well as the SSC emission of RS are suppressed). 
The predicted fluence of these sub-GeV flashes is high enough to be
detected by GLAST. 

Equally interestingly, the early photon-shock interaction leads to
interesting neutrino emission signatures. Besides the conventional
neutrino component powered by the photomeson interaction in the RS
region (DL01), in this paper we studied two new neutrino emission
components, i.e. a $\sim 5 \times 10^{15}$ eV component powered by
the photomeson interaction in the FS region, and a $\sim 10^{14}$ eV
component powered by photomeson interaction with the MeV photon flow. 
The detectability of the two components is comparable. The
spectrum of $10^{14}$eV neutrino is similar to that of the neutrinos
generated in the internal shock phase since the seed photons are
nearly the same, although the typical observed energy is softened by a
factor $\sim (\Gamma/\Gamma_0)^2$. The spectrum of the $5\times
10^{15}$eV neutrinos takes a more complicated form (see equation
[\ref{Spectrum1}]). 
A caveat is that these neutrinos overlap with the $\sim
5\times 10^{14}$eV internal shock neutrino-component (Waxman \&
Bahcall 1997), and hence, is difficult to distinguish. On the other
hand, they are generic regardless of whether the MeV emission is
powered by the internal shocks or not.

In this work, it is assumed that the wind medium is at rest at any
radius. This may not be the case if the radiation front of the GRB is
taken into account. As shown in Beloborodov (2002), due to the
large amount of $e^\pm$ pair loading (The pairs are created by the
interaction of the prompt $\gamma-$rays with the back-scattered
$\gamma$-rays by the wind medium), the medium will be accelerated and
the ejecta moves in a cavity until it reaches a radius $R_{\rm
gap}\sim 3.3\times 10^{15}E_{\gamma,53}^{1/2}$ (e.g., Beloborodov
2002). As a result, the shock crossing radius (see equation
(\ref{R_times})) should be 
$\sim R_{\rm gap}+R_\times({\rm wind})\sim 2 R_\times ({\rm
wind})$. The overlapping of the MeV photon flow and the shocked
regions is still significant since the ejecta essentially does not
decelerate at $R<R_{\rm gap}$. However, $f_\pi(1)$ and $f_\pi(2)$
involoved in the above calculation should be smaller by a factor of
about 1/4, so are the predicted neutrino fluxes. The optical depth for
the multi-GeV photons (see eq.[\ref{tau}]) is $\sim 1$, so that
they are still absorbed by the MeV photon flow and generate 
relativistic $e^\pm$ pairs, which re-scatter the initial MeV
$\gamma-$rays to the sub-GeV energy range. Notice that the
introduction of the radiation front will soften the FS shock emission
significantly since the upstream of the FS is $e^\pm$ rich. If we
assume that the number of the pairs is $k$ times that of the electrons
associated with protons, $\gamma_{\rm m}^{\rm f}$ (see
eq.[\ref{gamma_m}]) should be reduced to $\gamma_{\rm m}^{\rm 
f}/(1+2k)$. If $k$ is as large as several tens, the elecrons
accelerated in the FS will be cooled mainly via IC with the MeV photon
flow rather than synchrotron radiation and SSC emission. A significant
part of that IC component may be in sub-GeV energy range. Therefore
sub-GeV flashes are still expected when 
the radiation front effect is taken into account.

In this work, only the wind model is discussed. In the ISM model, if
the GRB is long enough, the overlapping of the 
MeV photon flow and the shocked regions could be also tight. In
this case, $R_\times ({\rm ISM})\sim 10^{17}$cm, and the optical depth
of GeV photons is very small (since $\tau_{\gamma \gamma}\propto
R^{-2}$, see eq.[\ref{tau}]). In such a case, GeV-TeV flashes
are expected, as suggested by Beloborodov (2005). Therefore, the
sub-GeV photon flashes are a signature for GRBs born in a 
stellar wind. 

We note that the overlapping effect increases the neutrino flux
predicted from the collapsar GRB model to be close to the one
predicted in the supranova model (cf. Dermer \& Atoyan 2003). 
This is because the overlapping MeV photons effectively provide a
convenient target for photomeson interactions, similar to the
external photons involved in the supranova model.  
  
\acknowledgments 
We thank Z. G. Dai for suggestions and the referee for
helpful comments. This work is supported by NASA NNG04GD51G and a NASA
Swift GI (Cycle 1) program (for B.Z.), the National Natural Science
Foundation (grants 10225314 and 10233010) of China, and the National
973 Project on Fundamental Researches of China (NKBRSF G19990754) (for
D.M.W.).

\end{document}